\documentclass[a4paper,10pt,twocolumn]{article}
\usepackage{amsmath}
\usepackage{amssymb}
\usepackage{amsfonts}
\usepackage{graphicx}
\usepackage{hyperref}

\newtheorem{theorem}{Theorem}
\newtheorem{definition}[theorem]{Definition}
\newtheorem{lemma}[theorem]{Lemma}

\begin{document}

\title{\textbf{A Review of Automated Formal Verification of Ad Hoc Routing Protocols for Wireless Sensor Networks}}
\author{
Zhe Chen\\
\emph{\normalsize College of Computer Science and Technology, Nanjing University of Aeronautics and Astronautics}\\
\emph{\normalsize 29 Yudao Street, 210016 Nanjing, Jiangsu, China}\\
\emph{\normalsize zhechen{@}nuaa.edu.cn}\\
Daqiang Zhang\\
\emph{\normalsize 1. School of Computer Science, Nanjing Normal University, 1 Wenyuan Road, 210046 Nanjing, Jiangsu, China}\\
\emph{\normalsize 2. Jiangsu Research Center of Information Security and Confidential Engineering}\\
\emph{\normalsize dqzhang{@}njnu.edu.cn}\\
Rongbo Zhu\\
\emph{\normalsize 1. College of Computer Science, South-Central University for Nationalities, Wuhan 430074, China}\\
\emph{\normalsize 2. The Bradley Department of Electrical and Computer Engineering, Virginia Tech, Blacksburg, VA 24061, USA}\\
\emph{\normalsize rongbozhu{@}gmail.com}\\
Yinxue Ma\\
\emph{\normalsize College of Computer Science and Technology, Nanjing University of Aeronautics and Astronautics}\\
\emph{\normalsize 29 Yudao Street, 210016 Nanjing, Jiangsu, China}\\
\emph{\normalsize yinxue.ma{@}gmail.com}\\
Ping Yin\\
\emph{\normalsize Science College, Jiangnan University, 1800 Lihu Avenue, Wuxi 214122, China}\\
\emph{\normalsize pyin{@}live.com} \\
Feng Xie\\
\emph{\normalsize Department of Computer Science, Shanghai Jiao Tong University, 800 Dongchuan Road, Shanghai 200240, China}\\
\emph{\normalsize fxie{@}sjtu.edu.cn}
}

\date{\normalsize (Received: 2011-11-01/Accepted: 2012-03-30)}

\twocolumn[
\begin{@twocolumnfalse}

Sensor Letters, Volume 11, Number 5, May 2013, pp. 752-764.

doi:10.1166/sl.2013.2653

\maketitle

\begin{abstract}
\noindent This paper surveys how formal verification can be used to prove the correctness of ad hoc routing protocols, which are fundamental infrastructure of wireless sensor networks. The existing techniques fall into two classes: verification on small-scale networks and verification on unbounded networks. The former one is always fully automatic and easy to use, thanks to the limited state space generated in verification. However, it cannot prove the correctness over all cases. The latter one can provide a complete proof based on abstractions of unbounded network. However, it usually needs user intervention and expertise in verification. The two kinds of technique are illustrated by verifications against some key properties such as stability, loop-freedom and deadlock-freedom. To conclude, they can be used to find faults and prove correctness, respectively. We believe that they can together aid the development of correct ad hoc routing protocols and their reliable implementations.

~

\noindent \textbf{Keywords}: formal verification, routing protocol, ad hoc network, model checking, theorem proving

~

\end{abstract}

\end{@twocolumnfalse}]

\section{Introduction}

A \emph{wireless sensor network} is composed of a large number of sensor nodes, which are densely deployed either inside the phenomenon to be supervised or very close to it \cite{ASSC02}. Wireless sensor networks are widely used for sensing, event detection etc., in military, environment, health, home and some commercial areas. These critical applications need correct and reliable behavior of their wireless sensor networks, because failures of expected functions can be catastrophic. Unfortunately, comparing to normal systems that are more stable and controllable, the correctness and reliability of these networks are much harder to ensure under variable topology, unpredictable dynamic environment and unbounded number of parallel processes.

Wireless sensor networks are usually organized as \emph{ad hoc network}, which is a kind of network such that each node participates in routing by forwarding data to other nodes, without relying on any centralized infrastructure such as router and base station \cite{Per01}. Nodes can join or leave at any time and are free to move around as they desire.

A \emph{routing protocol} is an algorithm used by nodes to determine a path for forwarding packets toward a destination. Ad hoc networks usually use distance vector routing, since centralized link state routing is not feasible due to its complexity and significant cost of propagating the link states. In distance vector routing protocols, each node maintains, for each destination, the name of next hop and the number of hops to reach the destination. Ad hoc routing protocols fall into two categories: \emph{table-driven} and \emph{on-demand} \cite{RT99}.

Routing protocol is one of the most fundamental infrastructures of ad hoc networks. High-level functions such as data transmission depend on the correct routes established by routing protocols. In other words, the correctness of routing protocols is a necessary condition of the reliable behavior of a network.

\subsection{From Simulation and Testing to Formal Verification}

The definition of correctness of routing protocols may consist of several aspects, such as stability, loop-freedom etc. The various key properties may be verified with different techniques, of which the two most important techniques are: traditional simulation and testing, and more recent formal verification.

\emph{Simulation} and \emph{testing} \cite{Mye79} are the most practiced techniques for verifying the correctness of routing protocols, since they are well understood by engineers and easy to use. The simulations of protocols can be automatically tested using computers (e.g. ns2), and the implementations of protocols can be tested on a real ad hoc network. Some faults can be discovered during simulation and testing, and then fixed by engineers. Thus, they can improve the probability that a protocol achieves the desired functions.

However, the most important shortcoming of simulation and testing is coverage. Note that we would first have to fix the network size and topology, then run the protocol for a limited length of time. A testing procedure starts from a given initial state, and only a single execution is observed and checked. Although the result is informative, it does not provide a complete verification, since a full proof should guarantee \emph{the correctness on all networks, over all lengths of time, under all possible initial states, and for every sequence of events that can occur}. Since simulation and testing cannot cover all these cases in running a protocol (and the coverage cannot be proved even if it is actually sufficient), they reduce the level of assurance in correctness. Indeed, there may be some subtle faults that remain hidden in a few tests, but might lead to serious failures in future operations.

\emph{Formal verification} (or formal method) is a kind of mathematical reasoning about systems. Formal verification provides more trustable verification results, thanks to their exhaustive searching of all possible executions. This means, if it is proved that a protocol satisfies some properties, then the result holds for all cases on all networks. The invention of some automated analysis techniques makes formal verification more and more feasible in practice. For instance, \emph{model checkers} \cite{CGP00,BK08} can automatically enumerate all possible executions of a system, and verify a property over these executions, while \emph{theorem provers} \cite{GM93} provide automated formal support for the creation and checking of proofs. However, model checkers suffer the state explosion problem\footnote{A \emph{state space} is a directed graph where each possible state of a dynamical system is represented by a vertex, and there is a directed edge if and only if there is a state transition from one state to another. In words, the state space consists of all the states generated in verification. The \emph{state explosion problem} is the fact that the state space grows exponentially in the number of processes, resulting in an intractable size.}, and the weakness of theorem provers is the significant user intervention. We believe that, the combination of manual mathematical reasoning, automated model checking and automated theorem proving is a more practical alternative for verifying large(-scale) systems.

\subsection{Difficulties in Formal Verification}

As we mentioned, the state explosion problem is one key impediment in making automatic model checking of large systems feasible in practice. Unfortunately, in the domain of verifying ad hoc routing protocols, this problem becomes more serious due to some new challenges.

Ad hoc routing protocols for mobile or wireless sensor networks are quite different from those for other types of network because of considerations like the large number of nodes, and highly variable and unreliable connectivity \cite{BOG02}. To address these new features, a routing protocol generally includes the following attributes:
\begin{enumerate}
  \item support an essentially unbounded number of replicated concurrent processes. Each process implements the protocol and runs on a node. The protocol should work well on unbounded number of nodes.
  \item dynamic and highly variable connectivity is assumed. Thus, the protocol should ensure reliable routing under continuously changing topology.
  \item unreliable connectivity is assumed which may lead to nondeterministic loss of packets. Thus, fault tolerance is required, i.e. the protocol should tolerate the loss of packets.
  \item real-time constraints are important, since many actions are triggered by timeout events.
  \item low-bandwidth links between nodes are assumed. Thus, the protocol should transmit packets and occupy bandwidth as less as possible.
\end{enumerate}
These attributes influence the way formal verification can be applied. As a result, verifying ad hoc routing protocols should consider the following special issues:
\begin{enumerate}
  \item A protocol should be verified on unbounded number of replicated processes that execute concurrently. A proof that works for two or three nodes is not trustable, unless it can be generalized. Therefore, the verification methodology should be optimized to deal with unbounded number of processes. For example, the optimizations of model checking should obviously reduce the generated state spaces.
  \item Nondeterministic connectivity changes should be modeled, in order to show the correctness under dynamic and highly variable connectivity.
  \item Nondeterministic loss of packets should be modeled, in order to show the correctness under unreliable connectivity.
  \item Some real-time features should be modeled and verified, in order to ensure the rationality of real-time constraints.
\end{enumerate}

It is easy to see that the \emph{unbounded number of processes} is one important source of the state explosion problem of verification, because they execute concurrently and lead to a huge state space. Since we must prove the correctness \emph{on all networks}, the complexity of verification is increased. Furthermore, model checking on all networks is generally not feasible, since we have infinite number of networks which cannot be enumerated.

Another source is the \emph{large number of possible topologies} for a given number of nodes in an ad hoc network. Generally, a network can be modeled as a graph. A node represents a wireless sensor or a mobile router etc. An edge represents the connection between two nodes. The graph (i.e. topology) may dynamically change due to variable links, mobility of nodes, etc. Note that each node is different, since they have unique identities. Suppose a bidirectional network containing $n$ nodes, a full graph contains $\frac{n \times (n-1)}{2}$ edges. So the number of possible topologies is
$$T = 2^\frac{n \times (n-1)}{2}$$
For unidirectional networks, this number grows to
$$T' = 2^{n \times (n-1)}$$
This is an $O(2^{n^2})$ order number of possible topologies\footnote{Note that the computation of possible topologies on Page 1182 of \cite{RA04} is not correct, since they imposed a lot of restrictions on possible topologies.}. For example, if $n=5$, $T = 1024$. When $n=6$, the number rapidly grows to $T = 32768$.

The large number of possible topologies influences the state space in two aspects. First, the number leads to a large number of possible initial states of the state space. Since we must prove the correctness \emph{under all possible initial states}, the complexity of verification is increased. Second, it is normal that the topology changes continuously during the operation, due to events such as broken or new links, broken or new nodes, etc. As a result, a large number of possible topologies may be nondeterministically reached during the execution. Since we must prove the correctness \emph{for every sequence of events that can occur}, the large number can greatly influence the complexity of verification.

Due to the $O(2^{n^2})$ order number of possible topologies, the state explosion problem becomes much more serious than verifying other applications. Indeed, in the literature, most current model checking tools can only be applied to networks containing five or six nodes. Memory can be rapidly ran out when checking protocols on larger networks.

\subsection{Overview}
We should consider two things when applying formal verification to ad hoc routing protocols. First, we need to specify the \emph{key properties} that should be satisfied by protocols, and then choose appropriate \emph{tools} to formally model systems and verify these properties. Second, we should carefully use appropriate \emph{verification techniques} to deal with the state explosion problem, and optimize our techniques with respect to two factors: the unbounded number of nodes and the large number of possible topologies for a given number of nodes.

This paper is organized as follows. The tools for verification are briefly introduced in Section \ref{Sec:tool}. The key properties that are expected to satisfy are described in Section \ref{Sec:prop}. The two classes of formal verification techniques are reviewed in Section \ref{Sec:tech}. We discuss challenges and future work in Section \ref{Sec:future}, and conclude in Section \ref{Sec:conc}.

The case studies in this paper concern the Border Gateway Protocol (BGP) \cite{RL95,RL06} which is the principal exterior gateway protocol, and the protocols for interior gateway such as the Routing Information Protocol (RIP) \cite{Hed88,Mal93,Mal94} which is a table-driven distance vector routing protocol using the asynchronous distributed Bellman-Ford protocol, the Ad hoc On-Demand Distance Vector (AODV) routing protocol \cite{PB99,PBD03}, and the Lightweight Underlay Network Ad-Hoc Routing Protocol (LUNAR) \cite{TGRW04} which is a simplified AODV-like protocol.

\section{Tool Support}
\label{Sec:tool}
In this section, we introduce some useful tools for verifying routing protocols. Some of the tools are fully automatic, e.g., SPIN and UPPAAL, whereas the others are based on user intervention, e.g., HOL.

The SPIN model checker \cite{Hol97,Hol03} has been widely used to verify communication protocols. A \emph{system} (e.g. a network) is modeled using the Promela language (a C-style language with communication operations), while a \emph{property} is specified as a finite state automaton or an LTL (Linear Temporal Logic \cite{HR04}) formula. SPIN can simulate the execution of the system, and more importantly, SPIN can perform an exhaustive search to check whether the property holds over all possible executions of the system. The verification process is fully automatic. If the property is violated in some execution, SPIN will provide a counterexample to aid fault diagnosis and revision of the system.

The UPPAAL model checker \cite{HDL04} is an integrated tool environment for modeling, validation and verification of real-time systems modeled as networks of timed automata, extended with data types (bounded integers, arrays, etc.) and channel synchronization. A \emph{system} is modeled using timed automata which are drawn in a graphical user interface, while a \emph{property} is specified as a CTL (Computation Tree Logic \cite{HR04}) formula. UPPAAL contains also a simulator and a checker. The verification process is fully automatic, and provides a counterexample if the property is violated in some execution.

The HOL theorem prover system \cite{GM93} is a general purpose verification environment based on Higher-Order Logic. A \emph{system} is modeled as functions using a functional programming language, while a \emph{property} is specified using higher-order logic. HOL uses a proof assistant that allows the user to construct proofs of such properties by using certain proof techniques. The proof assistant ensures the completeness and fault-freedom of the proofs. Theoretically, HOL is a general tool and capable of proving any mathematical theorem. Note that designing the proof strategy depends on expertise and user intervention, rather than complete automation.

These tools have their own merits and shortcomings. Model checkers, such as SPIN and UPPAAL, offer more comprehensive modeling language and fully automatic verification process, thus is easier to use. However, the bottlenecks of model checkers are \emph{memory} and \emph{expressiveness}. If we use them on a large network, the state space is so huge that the memory can be rapidly ran out. Furthermore, their models can only express limited number of processes, thus the unbounded number of nodes are generally beyond their expressiveness. On the other hand, theorem provers, such as HOL, offer more powerful but more complex mathematical infrastructures for developing more general proofs. Therefore, as a shortcoming, they need more \emph{expertise} and \emph{user intervention}, thus cost more \emph{programmer-months}.

We believe that a better strategy is to combine the two types of tool. We may first verify a protocol using automated model checkers such as SPIN and UPPAAL, on some typical, fixed and small(-scale) networks. Although the proof is not complete, some faults will be discovered and then fixed. Once we ensure the correctness of the protocol on these small networks, we may use theorem provers such as HOL to generalize the result.

For example, one instance of the strategy is to code a protocol first in SPIN and use HOL to address the limits of SPIN. This can be achieved by using HOL to prove the theorems on \emph{abstractions} showing properties such as: if properties $P_1,...,P_n$ hold for small sets of (maybe abstracted) nodes $S_1,...,S_n$, respectively, then property $P$ will hold for arbitrarily many nodes. We use the abstraction proofs in HOL to reduce the memory demands of SPIN proofs, since each property $P_i$ is checked on a small enough (abstract) network $S_i$. The abstraction proofs may also solve the problem of expressiveness. In fact, we largely reserve the comprehensibility and automation of model checking, but significantly reduce the memory demands of checking and increase the expressiveness of models by allowing user intervention in proving the theorems on abstractions. In essence, this strategy makes a tradeoff between memory consumption and the level of user intervention.

\section{Specifying Key Properties}
\label{Sec:prop}
Before verifying routing protocols, we need first to specify the properties that are expected to be satisfied. In this section, we introduce four key and generic properties to be verified and the way of specifying them. We start from defining two basic concepts.

First, we define the \emph{reachability} to a destination as follows.
\begin{definition}[Reachability]
A destination $d$ is \emph{reachable} from a node $n$, if there exists a path from $n$ to $d$ no greater than $N$ hops, where $N$ is the maximum hop count allowed by the protocol.
\end{definition}
In every routing protocol, the maximum number of hops in a route is always limited. For example, RIP sets the maximum number of hops to $N=15$. The protocols that use 16 or 32 bits for the hop count field of routing packet allow maximum $N=2^{16}-1$ or $N=2^{32}-1$ hops, respectively.

Second, the correctness of a routing protocol is defined as follows \cite{WPP04}.
\begin{definition}
\textbf{\emph{(Correct operation of an ad hoc routing protocol)}} If a destination node is reachable from a source node at one point in time, then the protocol must be able to find some path between the nodes. When a path has been found, and for the time it stays valid, it shall be possible to send packets along the path from the source node to the destination node.
\end{definition}
This generic definition implicitly indicates several concrete properties\footnote{Note that the correctness properties listed in \cite{Ole05} are not correct. For example, the property (1) in Section 5 of \cite{Ole05} cannot be satisfied, since it is normal that an ad hoc network is not fully connected at some time due to dynamic and variable connectivity. Furthermore, the author neither discussed how to express these properties using LTL or other formalisms.}. We list some important ones as follows.
\begin{enumerate}
  \item One important correctness property is \emph{stability}: for each destination $d$ that is \emph{reachable} from a node $n$, the node $n$ will eventually obtain a correct (optimal or shortest) route to $d$. A similar property is \emph{convergence}: a network can reach stability and does not change the established routes further, assuming the topology stays unchanged during this period.
  \item Another property related to stability is the \emph{real time convergence bound for stability}. We may want to calculate \emph{how much time} it must cost to reach stability, assuming the topology stays unchanged during this period of time.
  \item \emph{Loop-freedom} is an important property for distance vector routing protocols. A loop is a forwarding circle in the route from the source to its destination. If a loop route is established, packets will be transmitted in the loop but never reach the destination.
  \item It is also expected that \emph{deadlock} should not occur when running a protocol.
\end{enumerate}

Given a protocol, we need to show that these four generic properties are satisfied in all cases. These objectives can be achieved by formal verification, but may using different techniques and tools.

\section{Techniques for Formal Verification}
\label{Sec:tech}
In this section, we present different ways of modeling ad hoc networks and specifying key properties, and various techniques for verifying the models against the properties. We categorize these verification techniques into two classes.

The first class applies verification on small networks. Thanks to the limited size, the model can be established without abstractions. The advantage is that the system model is intuitive and easy to understand. Moreover, the verification could be fully automatic, thus much is gained in ease of use for non-experts, and some violations of properties could be found. These valuable information can be used to aid the correction of protocols. However, this method cannot cover all cases, since a network may consist of unbounded number of nodes.

The second class applies verification on unbounded networks. In this case, we try to show the correctness over all cases. Due to unbounded number of nodes, abstractions must be used in modeling. The advantage is that the proof is complete. However, due to the state explosion problem, a fully automatic checking is not possible. Thus, mathematical reasoning and abstract modeling must be involved. This needs high expertise in the verification domain.

No matter which class, we must first decide how to \emph{model} a network running ad hoc routing protocols by considering the following questions.
\begin{enumerate}
  \item How do we model each node running the protocol?
  \item How do we model the connectivity between nodes?
  \item How do we model communications such as broadcast and unicast?
  \item How do we model topology changes?
\end{enumerate}
Then we will consider the verification questions, such as, how do we specify the key properties using the formalism supported by the tools? We will show the answers by considering the two classes of verification techniques in the following subsections respectively.

\subsection{Verification on Small Networks}
Generally, a small network (e.g., three to six nodes) that runs a protocol on each node can be modeled and exhaustively checked with automated model checkers, since the state space is usually less than available memory. In this case, each node is explicitly modeled as a process, which is described using the Promela programming language of SPIN or formalized as a timed automaton in UPPAAL.

\subsubsection{Using SPIN}
In SPIN, a protocol is usually coded as a process type $proctype$ in Promela, which is then instantiated as several processes where each process denotes a node. They communicate through channels. Therefore, the most common verification methodology in the literature answers the modeling questions as follows:
\begin{enumerate}
  \item each node is modeled as a process, i.e., an instance of a process type that implements the protocol.
  \item the connectivity between nodes is either hard-coded into the process types, or managed by an intermediate connection process or a topology manager process.
  \item communications such as broadcast and unicast are modeled as messages through channels.
  \item the events of topology changes can be either nondeterministically generated by an intermediate environment process or a topology manager process, or implemented by nondeterministic choices in the process types.
\end{enumerate}
We will show the details in the following case studies.

The BGP protocol was modeled using the SPIN model checker, and then the convergence property was verified on four sample configurations, including good, disagree, precarious and bad gadgets \cite{Huadmai11}. The convergence property was verified by using two LTL formulas. The checking results of the two formulas can together show whether a BGP routing policy configuration will converge on a solution for a particular network destination, and the probability of convergence, i.e. completely solvable, partially solvable or unsolvable.

In the model, (1) The routers are modeled as Promela processes. (2) The connectivity is hard-coded into the process types through establishing communication channels. (3) Communications are modeled as messages through channels. (4) The topology is assumed to be not changed during this period. The convergence property is decided by considering the following two properties, which are verified in SPIN over some small network models.
\begin{align*}
    <> [~]~p &\text{~~~~~~where $p$ denotes all channels are empty,}\\
             &\text{~~~~~~i.e., possibly converge.}\\
    <> [~]~q &\text{~~~~~~where $q$ denotes some channels are not empty,}\\
             &\text{~~~~~~i.e., possibly never converge.}
\end{align*}
In words, the first formula specifies that the protocol will possibly converge in the future, i.e., formally, after some time in the future of any execution, for all the succeeding states, it always holds that all channels are empty. Similarly, the second formula specifies that the protocol will possibly never converge in the future.

The results of the verifications of the two formulas can help draw a conclusion whether a BGP network configuration will guarantee convergence for a particular network destination. Note that it is hard to generalize this approach to large networks due to the state explosion problem, and also dynamic topology changes are not addressed in the work.

\begin{figure}
  \centering
  \includegraphics[scale=1]{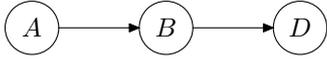}\\
  \caption{A Simple 3-Node Network}\label{Fig:3NodeNet}
\end{figure}

A model of AODV protocol was created using Promela in SPIN (302 lines of code), and it was attempted to prove loop-freedom for a simple 3-node network \cite{BOG02}\footnote{All the codes used in \cite{BOG02} are available at \url{http://www.cis.upenn.edu/verinet/RoutingVerification}}. The network consists of three nodes, A, B and D, as shown in Fig. \ref{Fig:3NodeNet}.
The node A has an active link to B, and B has an active link to D. Both A and B want to send data to the only destination D. The link B-D is fragile and may be broken at any time.

In the model, (1) The process type of AODV protocol is instantiated as three processes running at the three nodes. The timeout events are modeled as nondeterministic events that can occur at any time (in fact, it is an over-approximation), since Promela has no notion of real time. (2) The connectivity of the topology is hard-coded as a connection process running in parallel that captures all sent messages and delivers them when appropriate. (3) Broadcasting messages are captured by the process, which then delivers these messages to the sender's neighbors immediately. For such a fixed small topology, hard-coding has significant performance advantages in SPIN. (4) The fragility of the B-D link is modeled as an environment process which can nondeterministically send link-broken events to both B and D at any time. The property to verify is that there cannot be a routing loop between A and B, for any sequence of events. The following LTL formula expresses the loop-free property:
$$[~]~(!(next_D(A)==B~\&\&~next_D(B)==A))$$
In words, for all the states in any execution, it is should never occur that the next pointers of A and B for the destination D point to each other.

By running SPIN, a number of counterexamples are found and guide the revision of the draft of the protocol (version 2). The design and model of the protocol are modified several times until SPIN cannot find a loop. Note that SPIN cannot be sure that the fixed protocol will produce no loops for any network, since we only considered a 3-node network rather than unbounded nodes. However, it may indicate that we have found a possibly loop-free solution.

It is also suggested to model each node as a process in \cite{Ole05}. These processes manage their own connectivity to other nodes, and also the communication functions. Furthermore, a Location Manager (LM) process is specified to generate topology changes. That is, it nondeterministically decides when and what sensors change their locations and their connectivity to other nodes. The location manager process supports the following activities\footnote{Note that this definition of Location Manager is slightly different from \cite{Ole05}, since we also take into account the variable edges.}:
\begin{itemize}
  \item dynamic creation of new nodes, i.e., new processes.
  \item dynamic deletion of nodes, i.e., disconnection or termination of processes.
  \item dynamic creation of new edges, i.e., new connection.
  \item dynamic deletion of edges, i.e., disconnection between two nodes.
\end{itemize}
Node processes listen to LM, and adjust their connectivity with other nodes based on received messages from LM. Note that this modeling approach can only be used on small networks containing finite nodes, due to the state explosion problem.

In some other works, it is also admitted that the state explosion problem is a key impedient of model checking routing protocols due to mobility of nodes. Therefore, people only considered and modeled a five node network including a sender, a receiver and three intermediates in \cite{RA04}. A lot of restrictions on the model configurations have been imposed to obtain a simplified \emph{basic model}, e.g., limiting the number of possible links of each node, constraints on the possible topologies, etc. These simplifications can significantly reduce the state space in checking, but may also greatly reduce the confidence of the result.

In the model, (1) Each node is modeled as a process of simplified protocol. (2) The connectivity is saved in the routing table of each node. (3) Each node X has a channel $braodcastX$ only through which it can receive messages. The broadcasting operation is coded into each process to send messages to connected neighbors. (4) Node mobility is realized by canceling all routing table entries of the moving node, and setting a new link initialization for this node. One shortcoming of this method is extensibility. If we want to extend the verification to the models containing more nodes, more efforts are required to modify the code.

Based on the established basic model, the properties of deadlock and loop-freedom are checked. However, as we mentioned, due to the limited number of nodes and the restrictions on the configurations (e.g., possible links and topologies), the verification result is not trustable enough. It is also suggested that, when checking on large networks, SPIN can be used in the supertrace mode with the bitstate hashing technique \cite{Hol98} to save memory, with the cost of slightly reducing the coverage to about 98\% \cite{RA04}.

A general model of LUNAR was created using Promela in SPIN (about 250 lines of code excluding comments) \cite{WPP04,WPP05}. In the model, (1) Each node is a process instantiating the process type of a simplified version of LUNAR, where communication ports are modeled as channels stored in an array indexed by node id. (2) Connectivity is modeled as a symmetric, two dimensional array of booleans. The matrix is symmetric since the protocol assumes bidirectional connections. (3) Broadcast is modeled by unicasting to all nodes with whom the sending node has connectivity at that time. The unicast operations consisting a broadcast are implemented atomically to ensure its atomicity. (4) Topology changes are modeled using a separate process, which can modify the global connectivity matrix at any time.

In the verification, only one source and one destination are selected. The topology and node transitions are also selected, so that the two nodes are always connected. A hop counter is defined to keep track of the number of nodes traversed by the packet before it reaches its destination.

Based on the model, the properties of deadlock and loop-freedom are checked. Loop-freedom is formalized as the assertion: a packet does not traverse more nodes than theoretically possible before the packet reaches its destination. Absence of deadlock is formalized as a liveness property such that the packet will eventually be delivered in the future.

Their first verification attempt is to generate nondeterministically a topology for a given number of nodes, then the protocol is verified on each generated topology. Thus, all possible topologies are considered. However, the state space is too large to be efficiently checked. When node mobility is added, an exhaustive search is no longer feasible even for a small number of nodes.

Therefore, their second attempt focuses on a few typical scenarios (containing four or five nodes, and once or twice simple topology changes) which seem to cause problems for the protocol. The bitstate hashing option is not used in the verification to ensure reliable result. The checker cannot exhaustively search all the states when there are more than five nodes due to the memory limit of 36GB.

\subsubsection{Using UPPAAL}
If we use UPPAAL, real time constraints could be expressed. For example, a UPPAAL model was constructed to check timing requirements of LUNAR \cite{WPP04}. The model uses more abstraction than the SPIN model, because more complex data structures than arrays are not available in UPPAAL. In the model,
\begin{enumerate}
  \item each node is modeled as a timed automaton implementing the protocol.
  \item connectivity is modeled as a symmetric, two dimensional array of booleans.
  \item broadcasting is handled similarly to unicasting, except the sender has to send a request to the medium automaton. The medium distributes the broadcasting to other automata. The connectivity check is deferred to the receiving nodes, and value passing is realized by a combination of global data holders and committed locations, since broadcast channels in UPPAAL can only be used for synchronization rather than value passing.
  \item topology changes are modeled using a separate automaton, which can modify the global connectivity matrix at any time.
\end{enumerate}
Again, in the verification, only one source node and one destination are selected. The source tries to set up a route to the destination after which it attempts to send a packet there. Different from the SPIN model, communication delays can be modeled here, in order to take into account the influence caused by the delay of transmitting packets between nodes, i.e., latencies of information flow.

In the verification, if the route could be set up, the initiating node \verb=lunar0= goes into the state \verb=unic_rrep_rec=. If the correct packet arrives at the receiver \verb=lunar1=, it goes into the state \verb=ip_rec_ok=. The properties can be expressed as CTL formulas:
\begin{itemize}
  \item deadlock freedom: \verb=A[] not deadlock=.
  \item route successfully set up: \verb=A<> lunar0.uni_rrep_rec=.
  \item IP packet delivered: \verb=A<> lunar1.ip_rec_ok=.
\end{itemize}
Their verification attempts applied the same constraints on initial topologies and topology changes as using SPIN, due to the state explosion problem.

\subsubsection{Summary}
Some faults of a protocol could be discovered in the automatic model checking on small networks. These tools provide useful counterexamples for debugging and revising the implementation of the protocol. However, the proof is not complete, i.e., it is not ensured that the properties will hold in all cases and on all networks. Anyway, the experiments on small networks provide a possibly correct solution. Thus, we need some techniques to generalize the results on unbounded network.

\subsection{Verification on Unbounded Networks}
Large networks containing unbounded nodes cannot be modeled and checked directly, because the state space is too large to be efficiently verified by automated model checkers. We may introduce theorem provers to obtain a complete proof based on some checked properties that hold on some small sets of (maybe abstracted) nodes.

\subsubsection{Stability}
Let us consider how to prove stability of a protocol on arbitrary network. We must show that, for each destination $d$ that is reachable from some nodes, the nodes will all eventually obtain a correct path to $d$. We denote by $N$ the maximum hop count allowed by the protocol.

Our case study will show the stability of RIP (here $N=15$). The proof starts from choosing an arbitrary destination $d$. This can simplify the verification to deal with only one destination.

An entry for a destination $d$ at a node $r$ consists of at least two parameters: (1) $hops(r)$ denotes the current estimate of the distance metric to $d$; (2) $NextR(r)$ denotes the next node on the route to $d$. Initially, the nodes connected to $d$ have their metric set to 1, while others have values greater than 1. The \emph{distance} (i.e. the length of shortest path) from $r$ to $d$ is defined as:
\begin{equation*}
D(r)= \left\{ \begin{aligned}
         &\text{$1$, if $r$ is connected to $d$,} \\
         &\text{$1+min\{D(s)~|~ s$ neighbor of $r\}$, otherwise.}
             \end{aligned} \right.
\end{equation*}
For $k \geq 1$, the \emph{$k$-circle} around $d$ is $C_k = \{r~|~D(r)\leq k\}$, i.e. the set of nodes whose distance to $d$ is no greater than $k$. For $1 \leq k \leq N$, the universe is \emph{$k$-stable} if the following properties both hold:
\begin{description}
  \item[(S1)] For every node $r \in C_k$, $hops(r)=D(r)$, i.e. every node $r \in C_k$ has its metric set to the actual distance. Moreover, if $r$ is not connected to $d$, then $D(nextR(r))=D(r)-1$, i.e. its next hop's distance to $d$ is one hop shorter, which ensures the correct direction of the link that leads to $d$.
  \item[(S2)] For every node $r \not\in C_k$, $hops(r)>k$.
\end{description}
In words, all nodes inside $C_k$ have converged to the correct routes, while the outer nodes may not have found their correct routes. The protocol aims to expand the $k$-stable circle until all nodes are contained in the $N$-stable circle.

Given a $k$-stable universe, a node $r$ such that $D(r)=k+1$ is \emph{$(k+1)$-stable} if $hops(r)=k+1$ and $nextR(r) \in C_k$. In words, a node $r$ at distance $k+1$ from $d$ has found a shortest route.

The stability of RIP, i.e. each node eventually finds all the shortest paths to its reachable destinations, is formally defined as follows.

\begin{theorem}[Stability of RIP]\label{BOG02:Thm6}
For any $k \leq N$, starting from an arbitrary initial state of the universe $\mathcal{U}$, for any fair sequence of update messages, there is a time $t_k$ such that $\mathcal{U}$ is $k$-stable at all times $t \geq t_k$.
\end{theorem}
In particular, we want to show that $N$-stability will be reached. The theorem can be decomposed into the following lemmas and proved by induction on $k$.
\begin{lemma}[Basis]\label{BOG02:Lem7}
The universe $\mathcal{U}$ is initially 1-stable.
\end{lemma}
\begin{lemma}[Preservation of Stability]\label{BOG02:Lem8}
For any $k\leq N$, if the universe $\mathcal{U}$ is $k$-stable at some time $t$, then it is $k$-stable at any time $t' \geq t$.
\end{lemma}
\begin{lemma}\label{BOG02:Lem9}
For any $k<N$ and node $r$ such that $D(r)=k+1$, if the universe is $k$-stable at some time $t_k$, then there is a time $t_{r,k} \geq t_k$ such that $r$ is $(k+1)$-stable at all times $t \geq t_{r,k}$.
\end{lemma}
\begin{lemma}[Progress]\label{BOG02:Lem10}
For any $k<N$, if the universe $\mathcal{U}$ is $k$-stable at some time $t_k$, then there is a time $t_{k+1} \geq t_k$ such that $\mathcal{U}$ is $(k+1)$-stable at all times $t \geq t_{k+1}$.
\end{lemma}

To prove these lemmas, RIP is modeled in both SPIN and HOL. The SPIN model of RIP has 141 lines of code, while the HOL model has 495 lines of code.

Lemma \ref{BOG02:Lem7} is easily proved by HOL. In fact, it is also easy to manually validate the lemma from the definition of $k$-stability.

To prove Lemma \ref{BOG02:Lem8}, we need to show that a $k$-stable universe remains $k$-stable after an arbitrary update message. This could be done with only HOL, or with HOL and SPIN together. If we use SPIN, we must construct a nice \emph{abstraction} of the universe to avoid the unbounded number of nodes and processes. The abstraction contains only three processes (207 lines of code).
\begin{enumerate}
  \item In a $k$-stable universe, the $k$-circle always sends the distance metric of $k$ hops to the outside world. Therefore, the $k$-circle can be modeled by a single process $P$ that always sends the distance metric of $k$ hops.
  \item All the nodes outside the $k$-circle send to the $k$-circle all the distance metrics strictly greater than $k$. Therefore, the outside world can be modeled by a process $Q$ that always sends arbitrary distances greater than $k$.
  \item For a node $r$ such that $D(r)=k+1$, it runs a normal RIP process $R$. We abstract its environment and replace it with processes $P$ and $Q$.
\end{enumerate}
Using this abstraction, the unbounded universe $\mathcal{U}$ effectively reduces to three processes $P$, $Q$, $R$. Note that the abstraction is \emph{finitary}, since it reduces the system to a fixed finite number of states. This makes the state space of the abstraction can be exhaustively explored by a model checker like SPIN. It is also \emph{property-preserving} with respect to the desired properties. Whenever the abstraction satisfies a property, the concrete system also satisfies the property. This makes sure that the verification over the abstraction is sound. We can prove in HOL that the abstraction is property-preserving for the node $r$.

Lemma \ref{BOG02:Lem9} is proved with SPIN using the same abstraction. The idea can be summarized as follows: we justify the abstraction using a theorem prover, and then we prove the property of the abstraction using a model checker.

Lemma \ref{BOG02:Lem10} is proved by HOL as an easy generalization of Lemma \ref{BOG02:Lem9}, because there are finite number of nodes. Finally, Theorem \ref{BOG02:Thm6} is proved by HOL from Lemmas \ref{BOG02:Lem7}, \ref{BOG02:Lem8}, \ref{BOG02:Lem10} by induction on $k$. Again, it is also easy to manually validate Lemma \ref{BOG02:Lem10} and Theorem \ref{BOG02:Thm6} instead of using HOL.

It is worth noting that the property of stability on networks of unbounded nodes is decomposed into several properties on some small abstractions consisting of only a few processes, which significantly reduces the memory consumption of model checking. Furthermore, note that the environment processes $P$ and $Q$ may generate some message sequences that are not possible in reality. This is a kind of property-preserving over-approximation in model checking. Additionally, sometimes automated theorem provers like HOL are not necessary for proving some lemmas or theorems. Because some proofs may be easily constructed by hand, especially when the proofs are either short or not complex.

\subsubsection{Real-time Convergence Bound for Stability}
Note that stability is proved under the assumption that the topology stays unchanged for some period of time. The \emph{real-time convergence bound} is how big that period of time must be. This bound is based on the following timing assumption: during every time interval of the length $\Delta$, each node gets at least one update message from each of its neighbors. For RIP, $\Delta$ equals 3 minutes.

We define the \emph{radius} of the universe with respect to $d$ as $R=max\{D(r)~|~r\text{ is a node}\}$, i.e. the maximum distance from $d$. The real-time convergence bound of RIP is the following one:
\begin{theorem}[Convergence Bound of RIP]
A universe of radius $R$ becomes $N$-stable within $min\{N,R\}\cdot \Delta$ time, assuming that there were no topology changes during that time interval.
\end{theorem}

Again, this theorem can be similarly decomposed into four lemmas and proved by induction on $k$. The verification methodology is similar to that of RIP stability discussed earlier. We will not repeat the details here, the reader is referred to Section 5 of \cite{BOG02}.

\subsubsection{Loop-freedom}
Let us consider how to prove loop-freedom of a protocol on arbitrary network. Our case study will show that the ADOV protocol is loop-free. As specified in its standard, AODV protocol uses sequence numbers to avoid the formation of routing loops, a well-known shortcoming of RIP.

For arbitrary node $n$ and destination $d$, we write $seqno_d(n)(t)$ to denote $n$'s sequence number for $d$ at the time $t$. Similar notation is used for $hops_d(n)(t)$ and $next_d(n)(t)$. The parameter $(t)$ may be omitted if the given time is clear. We write $restart(n)(t)$ to denote $n$ was restarted at time $t$.

Our objective is the following theorem.
\begin{theorem}[Loop-freedom of AODV]\label{BOG02:Lem16}
For an arbitrary network of nodes running AODV, there will be no routing loops formed.
\end{theorem}
Theorem \ref{BOG02:Lem16} can be inferred from the following theorem.
\begin{theorem}\label{BOG02:Lem17}
For every destination $d$, if $next_d(n)=n'$, then
\begin{enumerate}
  \item $seqno_d(n) \leq seqno_d(n')$, and
  \item if $seqno_d(n) = seqno_d(n')$, then $hops_d(n) > hops_d(n')$.
\end{enumerate}
\end{theorem}
The proof of Theorem \ref{BOG02:Lem16} by Theorem \ref{BOG02:Lem17} can be done in HOL. However, it is worth noting that HOL is not necessary. Because it is easy to construct a manual proof of Theorem \ref{BOG02:Lem16} assuming Theorem \ref{BOG02:Lem17}. Assume there is a loop $n_1 \rightarrow n_2 \rightarrow \cdots \rightarrow n_k \rightarrow n_1$, then we have $seqno_d(n_1) \leq seqno_d(n_2) \leq \cdots \leq seqno_d(n_k) \leq seqno_d(n_1)$ by Theorem \ref{BOG02:Lem17}(1). So, it is must be $seqno_d(n_1) = seqno_d(n_2) = \cdots = seqno_d(n_k) = seqno_d(n_1)$. By Theorem \ref{BOG02:Lem17}(2), we have $hops_d(n_1) > hops_d(n_2) > \cdots > hops_d(n_k) > hops_d(n_1)$. Note that $hops_d(n_1) > hops_d(n_1)$ is a contradiction. Therefore, there does not exist a loop.

Theorem \ref{BOG02:Lem17} can be decomposed into the following lemmas.
\begin{lemma}\label{BOG02:Lem18}
If $t_1 \leq t_2$, and $\forall t: t_1 < t \leq t_2$, $\neg restart(n)(t)$, then $seqno_d(n)(t_1) \leq seqno_d(n)(t_2)$.
\end{lemma}
\begin{lemma}\label{BOG02:Lem19}
If $t_1 \leq t_2$, and $\forall t: t_1 < t \leq t_2$, $\neg restart(n)(t)$, and $seqno_d(n)(t_1) = seqno_d(n)(t_2)$, then $hops_d(n)(t_1) \geq hops_d(n)(t_2)$.
\end{lemma}
\begin{lemma}\label{BOG02:Lem20}
If $next_d(n)(t)=n'$, then there exists a last update time $lut \leq t$, such that:
\begin{enumerate}
  \item $seqno_d(n)(t) = seqno_d(n')(lut)$, and
  \item $hops_d(n)(t) = 1 + hops_d(n')(lut)$, and
  \item $\forall t': lut <t' \leq t$, $\neg restart(n')(t')$.
\end{enumerate}
\end{lemma}

Lemma \ref{BOG02:Lem18} says that the sequence number for a single destination never decreases over time, unless the node restarts. Lemma \ref{BOG02:Lem19} says that the hop count does not increase over time if the sequence number does not change. Lemma \ref{BOG02:Lem20} says that if $n$ points to $n'$, then this must be the result of an update message sent from $n'$ to $n$ at time $lut$, and $n'$ cannot have restarted after time $lut$. It is easy to see that the lemmas together imply Theorem \ref{BOG02:Lem17} by constructing a hand proof.

Now we consider how to prove the three lemmas. Note that Lemmas \ref{BOG02:Lem18} and \ref{BOG02:Lem19} are properties over a single node $n$, and Lemma \ref{BOG02:Lem20} is a property over two nodes $n$ and $n'$. This means, we have reduced the loop-free property on an unbounded network to several local properties on one or two processes. Each of the lemmas is individually proved in SPIN.

At first, we model AODV protocol in SPIN. The model has 302 lines of code.

To prove Lemma \ref{BOG02:Lem18}, we construct an \emph{abstraction} of the universe to avoid the unbounded number of nodes and processes. The abstraction contains only two processes.
\begin{enumerate}
  \item The node $n$ runs a normal AODV process $A$.
  \item All the nodes other than $n$ are modeled as a single environment process $E$ that generates all possible messages to $A$.
\end{enumerate}
Using this abstraction, the unbounded universe is effectively reduced to two processes $A$ and $E$. We need to prove that, in this model, the sequence number of $A$ never decreases. The abstraction is finitary and property-preserving. This makes the abstraction can be automatically checked in SPIN.

Lemmas \ref{BOG02:Lem19} and \ref{BOG02:Lem20} are proved with SPIN using the same abstraction. Each of the lemmas can be effectively checked in SPIN, thanks to the small number of processes. Finally, Theorem \ref{BOG02:Lem17} is easily proved by Lemmas \ref{BOG02:Lem18}, \ref{BOG02:Lem19}, \ref{BOG02:Lem20} with HOL or hand proof.

It is worth noting that the property of loop-freedom on networks of unbounded nodes is decomposed into several local properties on a small abstraction consisting of only two processes, which significantly reduces the memory consumption of model checking. Furthermore, note that the environment process $E$ may generate some message sequences that are not possible in reality. This is a kind of property-preserving over-approximation in model checking. Additionally, sometimes automated theorem provers like HOL are not necessary for proving some lemmas or theorems. Because some proofs may be easily constructed by hand, especially when the proofs are either short or not complex.

\section{Challenges and Future Work}
\label{Sec:future}
As we know, the biggest challenge comes from the model checking algorithms, which generate a huge and intractable state space in checking. To address this issue, new improvements of checking algorithms are being developed to reduce the state space, such as \cite{CM10b}. And some parallel model checking algorithms have been proposed recently \cite{BHV00,BBS01}. Extending existing verification tools with distributed checking algorithms may gain higher speed and better performance. This may enable the automatic checking of larger fixed networks and a wider range of scenarios.

Another challenge could be enhancing the automation level of theorem provers. Currently, most engineers are not capable of using provers, due to the lack of enough mathematical knowledge on that. Less user interaction may accelerate the application of theorem provers in industrial verification practice.

Challenge also comes from the engineers' expectation that the verification should be done on the real implementations of protocols such as codes, rather than on the models. As we know, subtle but fatal faults may be introduced, even when we are implementing a verified correct model. Direct checking of real codes would be more attractive and useful to engineers, since it can fill the gap between the model and the implementation of a protocol.

In the future, we may study how to assure the security of routing protocols, which is related to but different from the correctness and reliability of routing protocols that have been discussed in this survey. Some recent works tried to explore the approaches for model checking wireless sensor network security protocols, in order to decide the existence of attacks, such as route disruption, route diversion etc. \cite{TCCDC09,BMV10}.

Another future direction could be the development of the languages for constructing wireless sensor network applications. A recent work introduced such a language named \emph{Insense}, which supports a component-based model for wireless sensor network applications, and can facilitate the construction of low-level software for a concurrent, real-time and resource-constrained computing environment \cite{DBLM08}. Furthermore, the correctness of the implementations of \emph{send} and \emph{receive} operations was verified using SPIN on some typical small network models consisting of less than four nodes \cite{SLMDBMS09}.

Finally, we may also explore other techniques for verifying protocols such as runtime monitoring and control \cite{CM10a,CM09d}, and their theoretical foundations like $\omega$-automata \cite{Chen11}. We will try to apply these techniques to real-world applications like RFID networks \cite{ZZG11} and pervasive computing environments \cite{ZCH11,ZZL12}.

\section{Conclusion}
\label{Sec:conc}
In this paper, we reviewed two classes of formal verification techniques for verifying the correctness of ad hoc routing protocols: verification on small-scale networks and verification on unbounded networks. The former one is always fully automatic and easy to use, thanks to the limited state space generated in verification. However, it cannot prove the correctness over all cases. The latter one can provide a complete proof based on abstractions of unbounded network. However, it usually needs user intervention and expertise in verification. The two kinds of technique are illustrated by verifications against some key properties such as stability, loop-freedom and deadlock-freedom. To conclude, they can be used to find faults and prove correctness, respectively. It has been shown that these verification techniques for key properties are feasible in practice. We believe that they can be applied to other protocols of Internet Engineering Task Force (IETF), and together aid the development of correct ad hoc routing protocols and their reliable implementations.

\section*{Acknowledgement}
This work was supported by the National Natural Science Foundation of China (61100034, 61170043, 61103185), the China Postdoctoral Science Foundation (20110491411), the Jiangsu Planned Projects for Postdoctoral Research Funds (1101092C), the Start-up Foundation of Nanjing Normal University (2011119XGQ0072), the Natural Science Foundation of the Higher Education Institutions of Jiangsu Province (11KJB520009), and the Major Program of Natural Science Foundation of Jiangsu Province (BK211005). The authors wish to thank the editor and the anonymous referees for their detailed comments and helpful suggestions.

\bibliographystyle{plain}
\bibliography{bib_wanet,bib_veri,bib_us}

\end{document}